\documentclass[lettersize,journal]{IEEEtran}

\usepackage{amsmath,amsfonts}
\usepackage{algorithmic}
\usepackage{algorithm}
\usepackage{array}
\usepackage[caption=false,font=normalsize,labelfont=sf,textfont=sf]{subfig}
\usepackage{textcomp}
\usepackage{stfloats}
\usepackage{url}
\usepackage{verbatim}
\usepackage{graphicx}
\usepackage{cite}
\usepackage{color}
\begin{document}


\title{Effective DoF-Oriented Optimal Antenna Spacing in Near-Field XL-MIMO Systems }
\IEEEoverridecommandlockouts

\author{Xianzhe~Chen,
        Hong~Ren,~\emph{Member},~\emph{IEEE},
        Cunhua~Pan,~\emph{Senior~Member},~\emph{IEEE},
        Zhangjie~Peng,
        and~Jiangzhou~Wang,~\emph{Fellow},~IEEE

\vspace{-0.9cm}

\thanks{\emph{(Corresponding author: Cunhua Pan)}}

\thanks{X. Chen, H. Ren, C. Pan and Jiangzhou Wang are with the National Mobile Communications Research Laboratory, Southeast University, Nanjing 210096, China (e-mail: chen.xianzhe, hren, cpan, j.z.wang@seu.edu.cn).}

\thanks{Z. Peng is with the College of Information, Mechanical and Electrical Engineering, Shanghai Normal University, Shanghai 200234, China (e-mail: pengzhangjie@shnu.edu.cn).}

}

\maketitle

\newtheorem{lemma}{Lemma}
\newtheorem{theorem}{Theorem}
\newtheorem{remark}{Remark}
\newtheorem{corollary}{Corollary}
\newtheorem{proposition}{Proposition}

\begin{abstract}
This letter investigates the optimal antenna spacing for a near-field XL-MIMO communication system
from the perspective of the array gain.
Specifically, using the Green's function-based channel model, the letter analyzes the channel capacity, which is related to the effective degrees-of-freedom (EDoF). 
Then, the letter further investigates the applicability of two EDoF estimation methods.
To increase EDoF, this letter focuses on analyzing the impact of antenna spacing.
Furthermore, from the perspective of the array gain, the letter derives an approximate closed-form expression of the optimal antenna spacing,
at which EDoF is maximized and 
the array gain at the antenna nearest to the focused antenna of the transmit array becomes zero.
Finally, numerical results verify the main results of this letter.


\end{abstract}

\vspace{-0.3cm}

\section{Introduction}

In recent years, extremely large-scale multiple-input multiple-output (XL-MIMO) has drawn extensive research attention to meet the requirements of 
ultra-reliability, high capacity densities, extremely low-latency and low-energy consumption in future sixth-generation (6G) communications \cite{BJORNSON20193,8766143}.
Compared to the conventional massive MIMO, XL-MIMO deploys antennas with a high order-of-magnitude in the number to achieve
extremely high spectral efficiency.
The increased antenna number not only enlarges the array size, 
but also pushes the system's operation environment from the conventional far-field region to the near-field region.
As a result, new channel characteristics appears such as the spherical wavefront, spatial non-stationarity and so on \cite{10098681,10496996}.

Researchers have shown great interests of the benefits brought by the spherical wavefront in near-field XL-MIMO systems. 
For the conventional far-field channels, the degrees-of-freedom (DoF) of the line-of-sight (LoS) path is extremely limited, due to the single spatial angle of the planar wavefront.
On the contrary, in the near-field scenario, where the spherical wavefront is taken into consideration, the spatial angles varying over the whole transmit/receive array bring a great increase in DoF, thus largely improving the channel capacity.

Therefore, it is crucial to investigate the DoF in near-field XL-MIMO systems.
Meanwhile, since the channel capacity depends mostly on the orthogonal sub-channels with significant singular values,
researchers have paid extensive attention to the effective DoF (EDoF), i.e., the number of significant singular values of the channel matrix.
In \cite{2019Waves}, an estimation on EDoF was proposed with a concise expression from the perspective of the maximum number of intensity fringes.
Based on two-dimensional (2D) sampling theory arguments, approximate expressions were derived in \cite{9139337} for the EDoF of communication channels between a large intelligent surface (LIS) and a small intelligent surface (SIS). 
In \cite{9860745}, the authors investigated the EDoF of the systems with two non-parallel arrays and analyzed its impact on the channel capacity.
In \cite{yue2024}, directivity-aware EDoF in XL-MIMO systems was analyzed.
For a free space MIMO system, EDoF was studied in \cite{9650519}, which could be calculated with the channel matrix.
In \cite{wang2024}, the authors derived closed-form EDoF expressions with Green’s function-based channels in near-field XL-MIMO systems.

According to the aforementioned literature, 
the channel capacity of XL-MIMO systems can be largely improved by increasing the EDoF.
By increasing the number of antennas \cite{9650519} or decreasing the distance between the transmit and receive arrays \cite{9860745},
the EDoF can be increased.
However, the increase of EDoF in these ways is very limited,
which means large EDoF can only be achieved when an excessive number of antennas is deployed or the distance between the transmit and receive arrays is very close,
which however is not practical.
In order to increase the EDoF effectively, our work focuses on the impact of antenna spacing on EDoF.
From the perspective of the array gain, we investigate the relationship among the antenna spacing, EDoF and array gain.
The main contributions of this letter are summarized as follows:
a) Considering a near-field XL-MIMO communication system, we find that there exists a threshold of antenna spacing, before which the EDoF increases fast with the antenna spacing. After that, the EDoF begins to drop off.
b) We derive the closed-form expression of the threshold from the perspective of the array gain. 
c) The numerical results verify the relationship among the antenna spacing, EDoF and array gain. They also show that the derived threshold is the boundary within which two widely used EDoF estimation methods are effective.

\vspace{-0.3cm}

\section{System Model}

\begin{figure}[t]
\vspace{-0.5cm}
  \centering
  \includegraphics[scale=0.45]{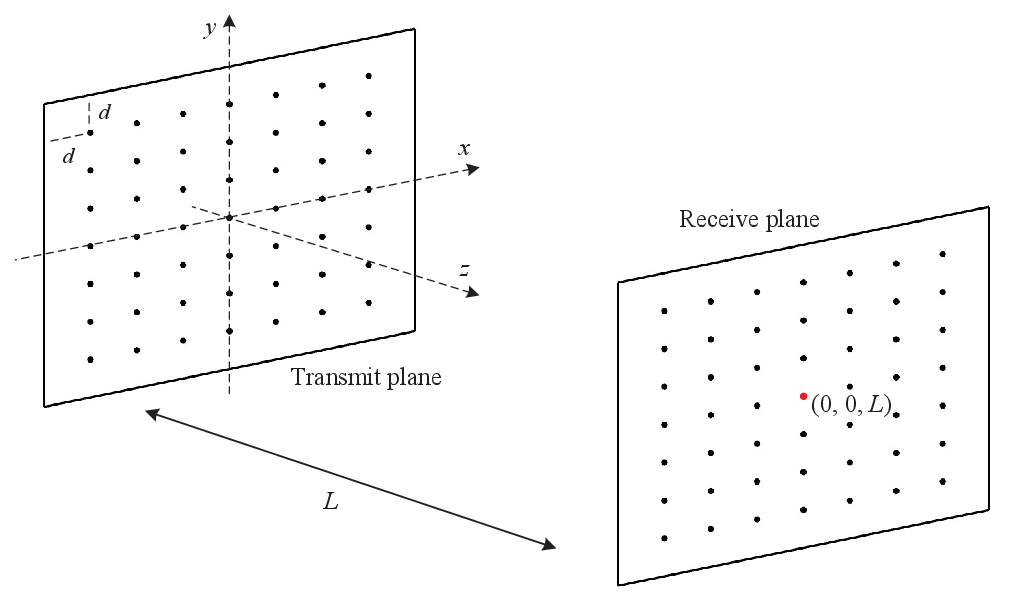}\\  \vspace{-0.2cm}
  \caption{System Model}\label{p1} \vspace{-0.4cm}
\end{figure}

Consider a near-field XL-MIMO communication system as depicted in Fig. \ref{p1}, where $N_{\mathrm{S}}$ point antennas in the transmit plane send signals to $N_{\mathrm{R}}$ point antennas in the receive plane.
For a monochromatic source $\psi \left( \mathbf{r} \right) $ at the transmit plane, the resulting Helmholtz wave equation is given by \cite{2019Waves} \vspace{-0.15cm}
\begin{equation}\label{eq1} \vspace{-0.2cm}
  \nabla ^2\phi \left( \mathbf{r} \right) +k^2\phi \left( \mathbf{r} \right) =\psi \left( \mathbf{r} \right), 
\end{equation}
where $k$ is the wave number, and $\phi \left( \mathbf{r} \right)$ represents the generated wave.
Then, the corresponding Green's function is obtained as \vspace{-0.2cm}
\begin{equation}\label{eq2}  \vspace{-0.2cm}
  \mathrm{G}\left( \mathbf{r};\mathbf{r}\prime \right) =-\frac{\exp \left( ik\left| \mathbf{r}-\mathbf{r}\prime \right| \right)}{4\pi \left| \mathbf{r}-\mathbf{r}\prime \right|}.
\end{equation}
With the Green's function above, the generated wave at the receive plane is expressed as \vspace{-0.2cm}
\begin{equation}\label{eq3} \vspace{-0.2cm}
  \phi \left( \mathbf{r}_{\mathrm{R}} \right) =\int_{S_{\mathrm{S}}}{\mathrm{G}\left( \mathbf{r}_{\mathrm{R}};\mathbf{r}_{\mathrm{S}} \right) \psi \left( \mathbf{r}_{\mathrm{S}} \right) d\mathbf{r}_{\mathrm{S}}}.
\end{equation}

For the considered system with $N_{\mathrm{S}}$ point transmit antennas (each at position $\mathbf{r}_{\mathrm{S}j}$, $j = 1,2,...,N_{\mathrm{S}}$), the generated wave at the point receive antenna $\mathbf{r}_{\mathrm{R}i}$ is the superposition of the electric fields from all the transmit antennas, which can be expressed as \vspace{-0.2cm}
\begin{equation}\label{eq4} \vspace{-0.2cm}
  \phi \left( \mathbf{r}_{\mathrm{R}i} \right) =\mathop {\Sigma}_{j=1}^{N_{\mathrm{S}}}\mathrm{G}\left( \mathbf{r}_{\mathrm{R}i};\mathbf{r}_{\mathrm{S}j} \right) s_j=\mathop {\Sigma}_{j=1}^{N_{\mathrm{S}}}g_{ij}s_j,
\end{equation}
where $s_i$ is the complex amplitude of the source at transmit antenna $\mathbf{r}_{\mathrm{S}j}$,
and $g_{ij}$ is defined as  \vspace{-0.1cm}
\begin{equation}\label{eq_gij}   \vspace{-0.1cm}
  g_{ij}=\mathrm{G}\left( \mathbf{r}_{\mathrm{R}i};\mathbf{r}_{\mathrm{S}j} \right) =-\frac{\exp \left( ik\left| \mathbf{r}_{\mathrm{R}i}-\mathbf{r}_{\mathrm{S}j} \right| \right)}{4\pi \left| \mathbf{r}_{\mathrm{R}i}-\mathbf{r}_{\mathrm{S}j} \right|}.
\end{equation}
Then, the received signals $\mathbf{f} \in \mathbb{C} ^{N_{\mathrm{R}} \times 1}$ at the receive antennas are given by \vspace{-0.1cm}
\begin{equation}\label{eq5}    \vspace{-0.1cm}
  \mathbf{f}=\mathbf{Gs}+\mathbf{n},
\end{equation}
where the vectors of the sources and the received signals are respectively given by \vspace{-0.1cm}
\begin{equation}\label{eq6} \vspace{-0.1cm}
  \mathbf{s}=\left[ s_1,s_2,...,s_{N_{\mathrm{S}}} \right] ^T, \;\;\;
  \mathbf{f}=\left[ f_1,f_2,...,f_{N_{\mathrm{R}}} \right] ^T.
\end{equation}
Vector $\mathbf{n}$ represents the additive white Gaussian noise (AWGN), with its elements following the distribution of 
${\cal C}{\cal N}\left(  {\bf{0}},{{\sigma }^2_{\mathrm{n}}}  \right)$. 
Matrix $\mathbf{G}$ is expressed as  \vspace{-0.1cm}
\begin{equation}\label{eq7}
  \mathbf{G}=\left[ 
  \begin{matrix}
	g_{11}&		g_{12}&		\cdots&		g_{1N_{\mathrm{S}}}\\
	g_{21}&		g_{22}&		\cdots&		g_{2N_{\mathrm{S}}}\\
	\vdots&		\vdots&		\cdots&		\vdots\\
	g_{N_{\mathrm{R}}1}&		g_{N_{\mathrm{R}}2}&		\cdots&		g_{N_{\mathrm{R}}N_{\mathrm{S}}}\\
  \end{matrix} \right]  ,
\end{equation}
which can be regarded as the channel matrix between the transmit plane and the receive plane.


\section{System Capacity} 

We first focus on the system capacity of the considered XL-MIMO system.
Assuming that each transmit antenna has equal transmit power under a total power of $P$,
the channel capacity can be calculated as \cite{1998On}   \vspace{-0.2cm}
\begin{equation*}
  C=\log _2\left( \det \left( \mathbf{I}_{N_{\mathrm{R}}}+\frac{P}{\sigma _{\mathrm{n}}^{2}N_{\mathrm{S}}}\mathbf{GG}^H \right) \right) 
\end{equation*}      \vspace{-0.2cm}
\begin{equation}\label{Req2}   \vspace{-0.2cm}
  \hspace{-1.2cm}
  =\sum_{i=1}^{n_{\mathrm{DoF}}}{\log _2\left( 1+\frac{P\mu _{i}^{2}}{\sigma _{\mathrm{n}}^{2}N_{\mathrm{S}}} \right)},
\end{equation}
where $n_{\mathrm{DoF}}$ is the DoF of the communication channel, and $\mu _{i}^{2}$ is the $i$-th largest eigenvalue of the channel correlation matrix $\mathbf{GG}^H$, which represents the channel gain of the corresponding orthogonal sub-channel. 


$n_{\mathrm{DoF}}$ in \eqref{Req2} equals the number of the non-zero eigenvalues of the correlation matrix.
Due to the large transmit and receive arrays in XL-MIMO systems, $n_{\mathrm{DoF}}$ is generally very high.
However, according to the capacity expression in \eqref{Req2}, the sub-channels with small eigenvalues only bring little contribution to the capacity. 
Thus, researchers are more interested in the significant eigenvalues, the number of which is defined as the EDoF, which is denoted by $n_{\mathrm{EDoF}}$.
Then, the capacity can be expressed as  \vspace{-0.1cm}
\begin{equation}\label{Req2-1}  \vspace{-0.15cm}
  C \approx \sum_{i=1}^{n_{\mathrm{EDoF}}}{\log _2\left( 1+\frac{P\mu _{i}^{2}}{\sigma _{\mathrm{n}}^{2}N_{\mathrm{S}}} \right)}.
\end{equation}

Intuitively, by performing singular value decomposition (SVD) of the channel matrix,
$n_{\mathrm{EDoF}}$ can be obtained by counting the number of the significant singular values.
In this work, we define the expression of $n_{\mathrm{EDoF}}$ as \vspace{-0.2cm}
\begin{equation}\label{eq_nEDoF} \vspace{-0.1cm}
 n_{\mathrm{EDoF}}=\underset{n}{\text{argmin}}\left\{ f\left( n \right) =\sum_{i=1}^n{\mu _{i}^{2}} \; \Bigg| \; \frac{f\left( n \right)}{\sum_{i=1}^{n_{\mathrm{DoF}}}{\mu _{i}^{2}}}  \geq 99.9\%  \right\}   .
\end{equation}
Though this direct solution gives a precise value of $n_{\mathrm{EDoF}}$,
it does not give an explicit closed form expression for $n_{\mathrm{EDoF}}$.
Besides, the SVD operation has a high computational complexity, e.g., ${\cal O} \left( N^3 \right)$ for an $N$-dimension matrix.

To address these issues, numerous methods are proposed to estimate $n_{\mathrm{EDoF}}$, among which two methods are widely used.
From the perspective of the maximum number of intensity fringes,
one way to estimate $n_{\mathrm{EDoF}}$ is given by \cite{2019Waves} \vspace{-0.1cm}
\begin{equation}\label{Req3} \vspace{-0.1cm}
  n_{\mathrm{EDoF1}}=\frac{A_{\mathrm{S}}A_{\mathrm{R}}}{\lambda ^2L^2},
\end{equation}
where $A_{\mathrm{S}}$ and $A_{\mathrm{R}}$ are the areas of the transmit and receive planes, respectively.
This method is suitable for the paraxial situation, 
where the separation of the transmit and receive arrays are relatively large compared to the apertures of them,
making the propagation waves almost parallel to the axis between the transmit and receive arrays.
The corresponding paraxial behavior means that the descending ordered singular values of the channel matrix are nearly constant before a relatively abrupt threshold, after which they drop off rapidly.

Another method to estimate $n_{\mathrm{EDoF}}$ is given by \cite{9650519} \vspace{-0.2cm}
\begin{equation}\label{Req4} \vspace{-0.1cm}
  n_{\mathrm{EDoF}2}=\frac{\mathrm{tr}^2\left( \mathbf{GG}^H \right)}{\left\| \mathbf{GG}^H \right\| _{\mathrm{F}}^{2}}=\frac{\left( \sum_i{\mu _{i}^{2}} \right) ^2}{\sum_i{\mu _{i}^{4}}}.
\end{equation}
The situation for which this method is suitable has not been explicitly given.
Here we provide a mathematical understanding for the applicability condition of \eqref{Req4}.

\begin{remark}\label{Re1}
Eq. \eqref{Req4} is also suitable for the paraxial situation.
Due to the paraxial behavior, the significant singular values of the channel matrix can be considered to be the same.
On the other hand, these significant singular values take up a dominant part of the total sum of singular values, since singular values drop off rapidly after the threshold, i.e., $n_{\mathrm{EDoF}}$.
Thus, the $n_{\mathrm{EDoF2}}$ in \eqref{Req4} can be approximated as \vspace{-0.15cm}
\begin{equation}\label{Req5} \vspace{-0.1cm}
  n_{\mathrm{EDoF}2}=\frac{\left( \sum_i{\mu _{i}^{2}} \right) ^2}{\sum_i{\mu _{i}^{4}}}\approx \frac{\left( n_{\mathrm{EDoF}}\mu ^2 \right) ^2}{n_{\mathrm{EDoF}}\mu ^4}=n_{\mathrm{EDoF}}.
\end{equation}
It is readily observed from \eqref{Req5} that as long as the XL-MIMO system works in paraxial situation,
$n_{\mathrm{EDoF2}}$ can well approximate $n_{\mathrm{EDoF}}$.
\end{remark}

\section{Impact of Antenna Spacing on EDoF}

\begin{figure}[t]
\vspace{-0.9cm}
  \centering
  \includegraphics[scale=0.4]{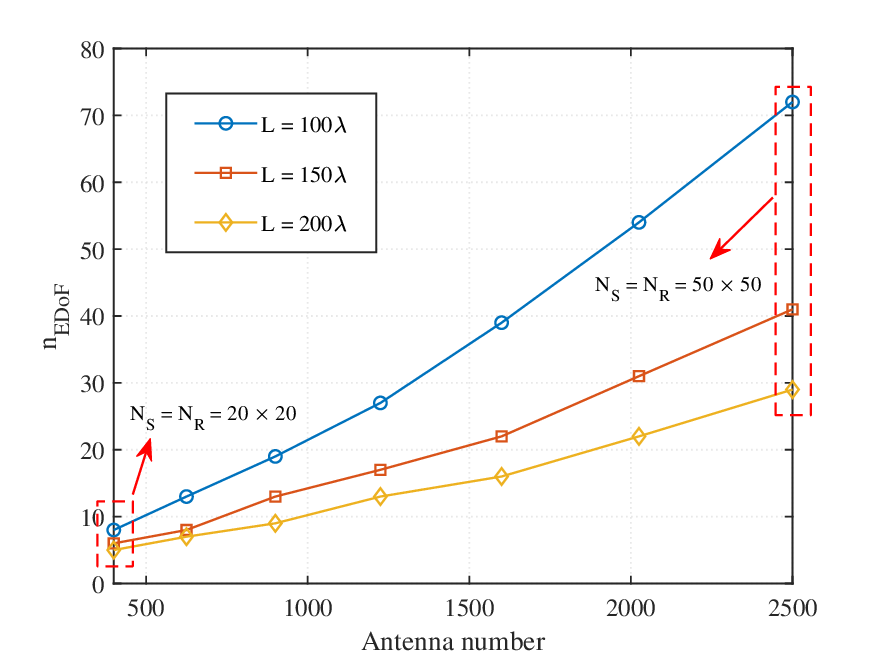}\\  \vspace{-0.2cm}
  \caption{The increase of $n_{\mathrm{EDoF}}$ with antenna number ($d = \frac{\lambda}{ 2} $)}\label{p2} \vspace{-0.4cm}
\end{figure}

Eq. \eqref{Req2-1} indicates that the channel capacity can be greatly improved by increasing $n_{\mathrm{EDoF}}$.
Intuitively, $n_{\mathrm{EDoF}}$ can be increased by adding more transmit and receive antennas.
In fact, the upper bound of $n_{\mathrm{EDoF}}$ is indeed largely increased in this way.
However, the increase of $n_{\mathrm{EDoF}}$ seems very limited compared to the increase of the number of antennas, which can be observed from Fig. \ref{p2}.
Besides, the excessive increase of antennas would bring unacceptable energy consumption and hardware cost.

\begin{figure}
  \centering
  \includegraphics[scale=0.4]{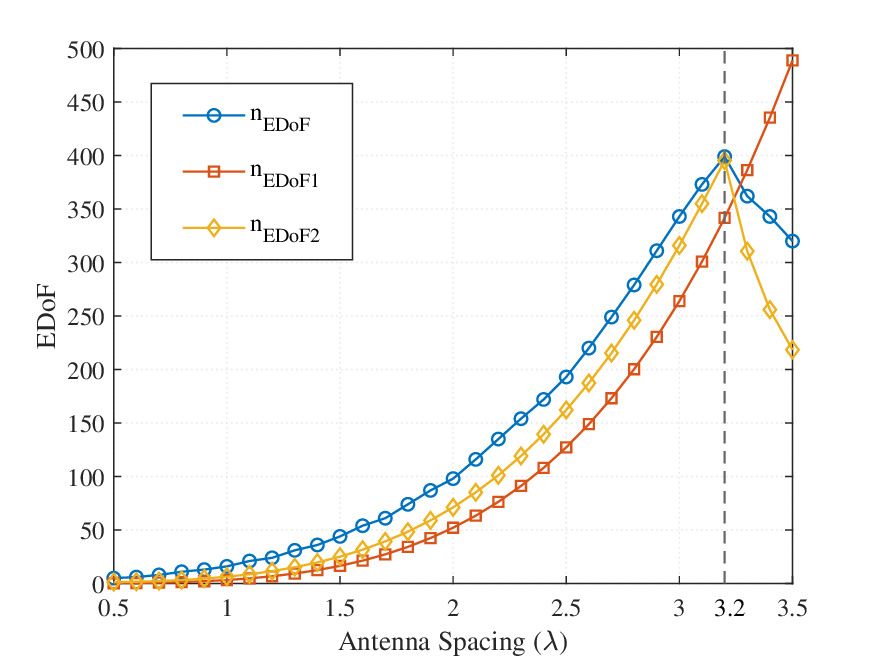}\\  \vspace{-0.2cm}
  \caption{The increase of $n_{\mathrm{EDoF}}$ with antenna spacing}\label{p3} \vspace{-0.4cm}
\end{figure}

To obtain a higher $n_{\mathrm{EDoF}}$ with given numbers of transmit and receive antennas,
we focus on the impact of the increase of antenna spacing on $n_{\mathrm{EDoF}}$.
It is observed from Fig. \ref{p3} that $n_{\mathrm{EDoF}}$ can be effectively increased by increasing the antenna spacing.
Moreover, it can be found that there is a threshold for antenna spacing, i.e., $d = 3.2\lambda$ in this simulation.
As the antenna spacing increases to the threshold, $n_{\mathrm{EDoF}}$ increases to its upper bound, i.e., 400 for the $20 \times 20$ transmit and receive planes. 
After the threshold, $n_{\mathrm{EDoF}}$ begins to drop off, and the two aforementioned methods to estimate $n_{\mathrm{EDoF}}$ begin to be invalid.
Therefore, this threshold is an important parameter for the XL-MIMO system design, and an approximate analytical expression of this threshold will be derived in the following.


\begin{figure}[t]
\vspace{-0.9cm}
  \centering
  \includegraphics[scale=0.45]{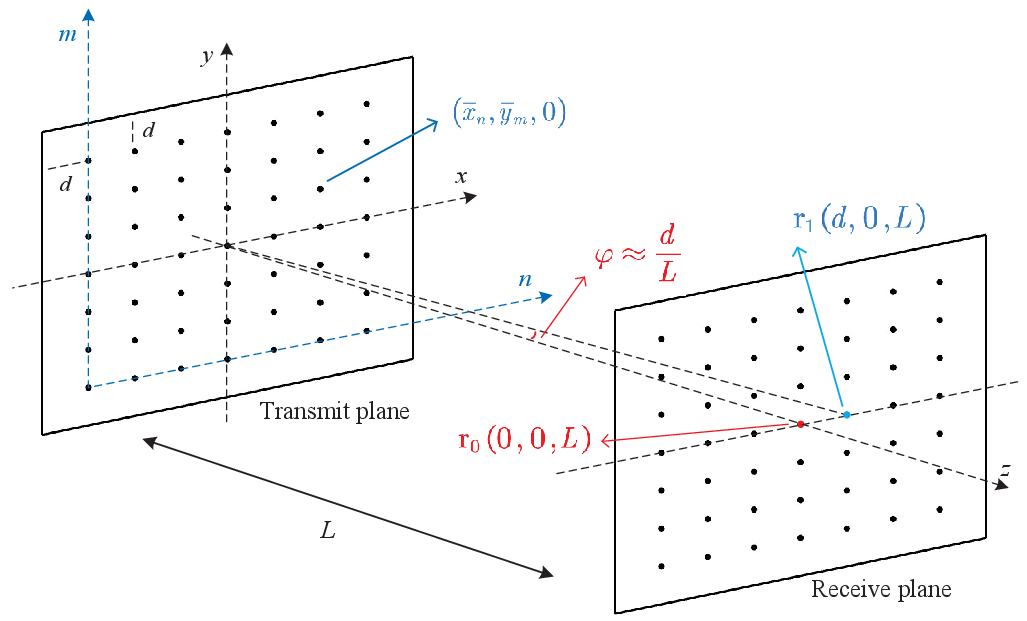}\\  \vspace{-0.2cm}
  \caption{System diagram for the derivation of the threshold}\label{p4} \vspace{-0.4cm}
\end{figure}

For ease of expression, we assume that the same size of uniform planar arrays (UPAs) is used in transmit and receive planes,
with the numbers of transmit and receive antennas set to $N_{\mathrm{S}}=N_{\mathrm{R}}=N$,
and the antenna spacing set to $d$.
Then, when the transmit array focuses on the antenna at position $\mathbf{r}_0 (0,0,L)$ in the receive plane, which is shown in Fig. \ref{p4},
the received signal at antenna $\mathbf{r}_0$ can be expressed as \vspace{-0.2cm}
\begin{equation}\label{ASeq1} \vspace{-0.2cm}
  f_0=\sum_{n=1}^{\sqrt{N}}{\sum_{m=1}^{\sqrt{N}}{t^0_{n,m}e^{i\theta _{n,m}^{0}}s}}+n,
\end{equation}
where $t^0_{n,m}$ represents the channel coefficient for antenna $\mathbf{r}_0$, which is a re-arranged version of $g_{ij}$ in \eqref{eq_gij} for ease of analysis, given by   \vspace{-0.3cm}
\begin{equation}\label{ASeq2} \vspace{-0.1cm}
  t^0_{n,m}=-\frac{\exp \left( ik\sqrt{\bar{x}_{n}^{2}+\bar{y}_{m}^{2}+L^2} \right)}{4\pi \sqrt{\bar{x}_{n}^{2}+\bar{y}_{m}^{2}+L^2}}\approx -\frac{e^{i\frac{2\pi}{\lambda}\sqrt{\bar{x}_{n}^{2}+\bar{y}_{m}^{2}+L^2}}}{4\pi L}.
\end{equation}
The approximation in \eqref{ASeq2} holds due to the fact that the power variations over arrays are negligible in radiative near field compared to the phase variations \cite{Towards6G}.
$\bar{x}_n$ and $\bar{y}_n$ are respectively given by  \vspace{-0.2cm}
\begin{equation}\label{ASeq3} \vspace{-0.25cm}
  \bar{x}_n=\left( n-\frac{\sqrt{N}+1}{2} \right) d, \;\;\;
  \bar{y}_m=\left( m-\frac{\sqrt{N}+1}{2} \right) d,
\end{equation}
where the index $\left( n,m \right) $ represents the relative position for antennas in the transmit plane, as shown in Fig. \ref{p4}.
And $\theta _{n,m}^{0}$ in Eq. \eqref{ASeq1} is the phase shift imposed on antenna $\left( n,m \right) $ in order to make the transmit array focus on $\mathbf{r}_0$.

From \eqref{ASeq1}, the signal-to-noise ratio (SNR) of the received signal at antenna $\mathbf{r}_0$ is expressed as  \vspace{-0.3cm}
\begin{equation}\label{ASeq4}  \vspace{-0.2cm}
  \mathrm{SNR}_0=\frac{P}{N\sigma _{\mathrm{n}}^{2}}\left| \sum_{n=1}^{\sqrt{N}}{\sum_{m=1}^{\sqrt{N}}{t^0_{n,m}e^{i\theta _{n,m}^{0}}}} \right|^2=\frac{P}{\sigma _{\mathrm{n}}^{2}}\frac{1}{\left( 4\pi L \right) ^2}\rho _0 ,
\end{equation}
where $\rho _0$ is the array gain at antenna $\mathbf{r}_0$ brought by the transmit array, which is defined as  \vspace{-0.3cm}
\begin{equation}\label{ASeq5}  \vspace{-0.2cm}
  \rho _0 \approx  \frac{1}{N}\left| \sum_{n=1}^{\sqrt{N}}{\sum_{m=1}^{\sqrt{N}}{e^{i\frac{2\pi}{\lambda}\sqrt{\bar{x}_{n}^{2}+\bar{y}_{m}^{2}+L^2}}e^{i\theta _{n,m}^{0}}}} \right|^2 .
\end{equation}
When the transmit array focuses on $\mathbf{r}_0$ in the receive plane, 
the array gain in Eq. \eqref{ASeq5} is maximized. 
Thus, $\theta _{n,m}^{0}$ is given by   \vspace{-0.2cm}
\begin{equation}\label{ASeq6} \vspace{-0.1cm}
  \theta _{n,m}^{0}=-\frac{2\pi}{\lambda}\sqrt{\bar{x}_{n}^{2}+\bar{y}_{m}^{2}+L^2}.
\end{equation}

We are interested in the case when $n_{\mathrm{EDoF}}$ is maximized. 
Since $n_{\mathrm{EDoF}}$ is related to the number of orthogonal sub-channels, 
it is reasonable to assume that when the transmit array focuses on one antenna, 
 $n_{\mathrm{EDoF}}$ is maximized when the interference imposed on the nearest antenna is minimized,
which means the array gain at the nearest antenna is minimized. 
To investigate the relationship between $n_{\mathrm{EDoF}}$ and the antenna spacing,
and obtain the threshold of the antenna spacing in Fig. \ref{p3}, 
we have the following theorem.

\begin{theorem}\label{T1}
When the transmit array focuses on antenna $\mathbf{r}_0 (0,0,L)$, 
the array gain at the nearest antenna $\mathbf{r}_1 (d,0,L)$ can be approximated as \vspace{-0.3 cm}
\begin{equation}\label{T1eq1} \vspace{-0.1cm}
  \rho _{1} \approx N\frac{\mathrm{sinc}^2\left( \frac{\sqrt{N}d^2}{\lambda L} \right)}{\mathrm{sinc}^2\left( \frac{d^2}{\lambda L} \right)}.
\end{equation}
When this array gain is minimized, i.e., $\rho _{1}$ becomes zero at the first time, 
$n_{\mathrm{EDoF}}$ is maximized.
In this case, 
we have $\frac{\sqrt{N}d^2}{\lambda L}=1$,
and thus we obtain the threshold of the antenna spacing given by \vspace{-0.2cm}
\begin{equation}\label{T1eq2}   \vspace{-0.1cm}
  d_{\mathrm{threshold}}=\sqrt{\frac{\lambda L}{\sqrt{N}}}.
\end{equation}

\end{theorem}

\begin{IEEEproof}
  Please refer to Appendix \ref{app_A}.
\end{IEEEproof}

Theorem \ref{T1} provides the threshold of the antenna spacing when $n_{\mathrm{EDoF}}$ is maximized given a number of antennas $N$,
which is crucial for the XL-MIMO system design.
Furthermore, more interesting insights can be found from Theorem \ref{T1}.
As indicated by Theorem \ref{T1}, $n_{\mathrm{EDoF}}$ approaches its maximum value when the system parameters satisfy  \vspace{-0.1cm}
\begin{equation}\label{T1eq3}  \vspace{-0.1cm}
  \frac{\sqrt{N}d^2}{\lambda L}   \approx  \epsilon \rightarrow  1.
\end{equation}
It is readily observed that $\epsilon$ increases only with the square root of the number of antennas $N$, but with the square of the antenna spacing $d$,
which verifies the discussion for Fig.~\ref{p2} and Fig.~\ref{p3} 
that $n_{\mathrm{EDoF}}$ increases fast with the increase of $d$, but slowly with the increase of $N$.
Besides, it can be found that $\epsilon$ is inversely proportional to the distance $L$ between the transmit and receive planes,
which means that $n_{\mathrm{EDoF}}$ decreases as $L$ increases.
This is consistent with the results of the existing literature \cite{9860745,9650519}.

\section{Numerical Results}  

\begin{figure}[t]
\vspace{-0.8cm}
  \centering
  \includegraphics[scale=0.4]{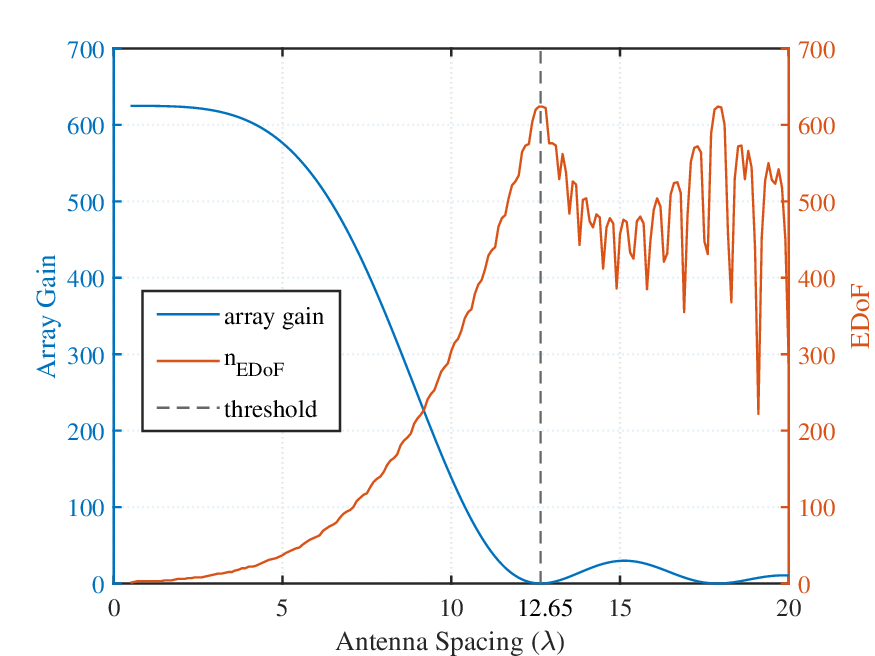}\\  \vspace{-0.2cm}
  \caption{Relationship between array gain and $n_{\mathrm{EDoF}}$}\label{p5} \vspace{-0.4cm}
\end{figure}

\begin{figure}[t]
  \centering
  \includegraphics[scale=0.4]{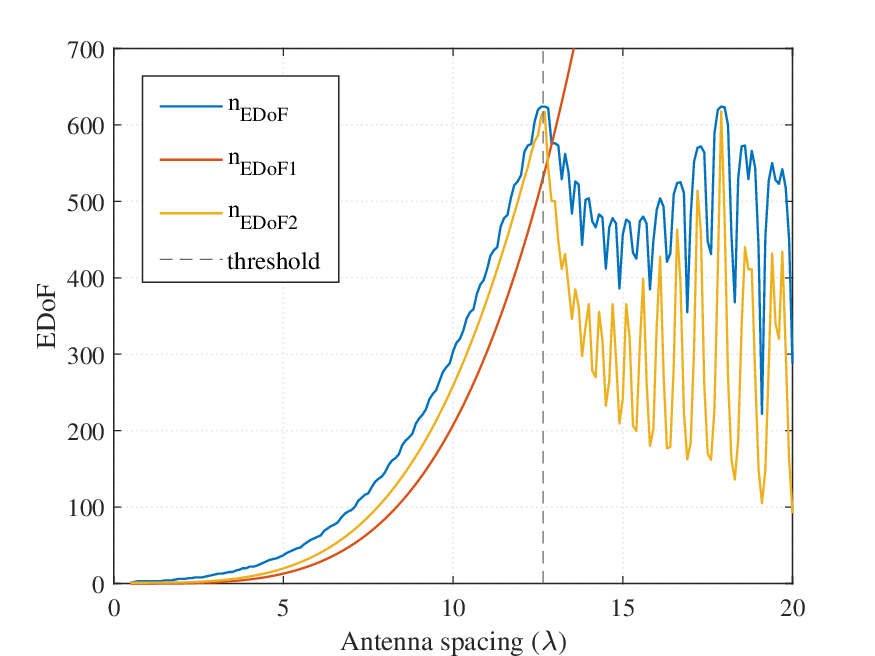}\\  \vspace{-0.2cm}
  \caption{Methods to estimate $n_{\mathrm{EDoF}}$}\label{p6} \vspace{-0.4cm}
\end{figure}

In this section, numerical results are provided to verify the main results of this letter and provide more insights.
We assume that the XL-MIMO system operates at 30 GHz, i.e., $\lambda = 0.01$ m.
UPAs for both transmit and receive planes are set to $N_{\mathrm{S}}=N_{\mathrm{R}}=N = 25 \times 25$, with the same antenna spacing $d$.
The distance between the transmit and receive planes is set to $L = 4000\lambda$.

Fig. \ref{p5} depicts the relationship between the array gain in \eqref{T1eq1} and $n_{\mathrm{EDoF}}$.
The blue line shows the array gain versus the antenna spacing $d$, 
while the orange line represents the $n_{\mathrm{EDoF}}$ versus $d$.
It is found that the array gain becomes zero at the threshold of antenna spacing $d_{\mathrm{threshold}} = 12.65\lambda $,
while $n_{\mathrm{EDoF}}$ increases to its peak, 
which verifies the accuracy of the results in Theorem \ref{T1}.
It can be noted that $n_{\mathrm{EDoF}}$ begins to drop when $d$ is slightly over the threshold,
which appears also in Fig. \ref{p3}.
This phenomenon can be well explained by Theorem \ref{T1}, 
since the array gain begins to increase due to its sinc-like characteristic.

Fig. \ref{p6} compares two widely used methods of EDoF estimation, which have been discussed in Section II. 
It is observed that both methods can well approximate $n_{\mathrm{EDoF}}$ when the antenna spacing is below the threshold.
However, after the threshold, both methods fail to approximate $n_{\mathrm{EDoF}}$.
This finding is valuable, as it shows the applicability condition of these two methods, 
which is when the antenna spacing is below the threshold given by \eqref{T1eq2}.

\begin{figure}[t]
\vspace{-0.8cm}
  \centering
  \includegraphics[scale=0.4]{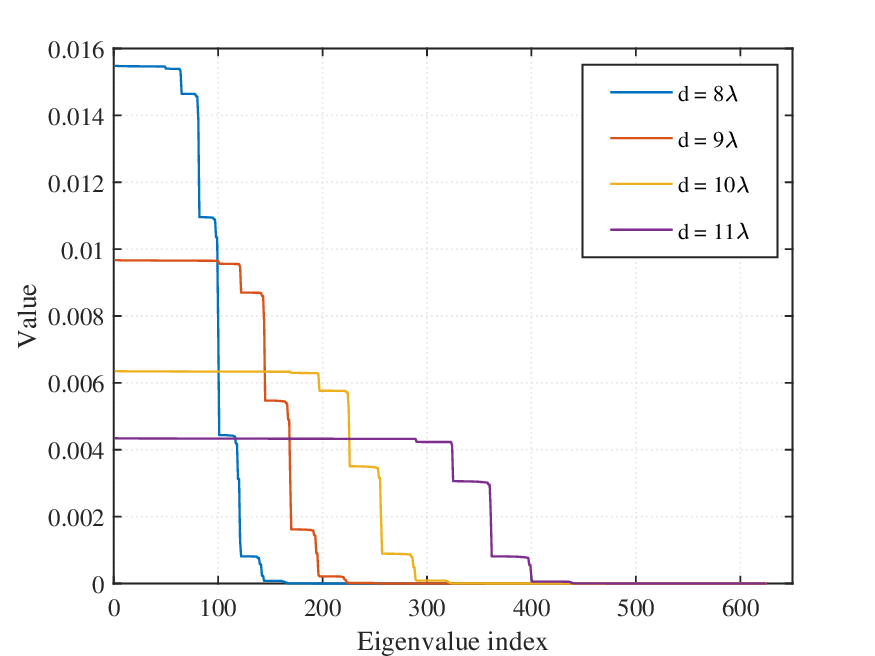}\\  \vspace{-0.2cm}
  \caption{Eigenvalues before the threshold}\label{p7} \vspace{-0.4cm}
\end{figure}

\begin{figure}[t]
  \centering
  \includegraphics[scale=0.4]{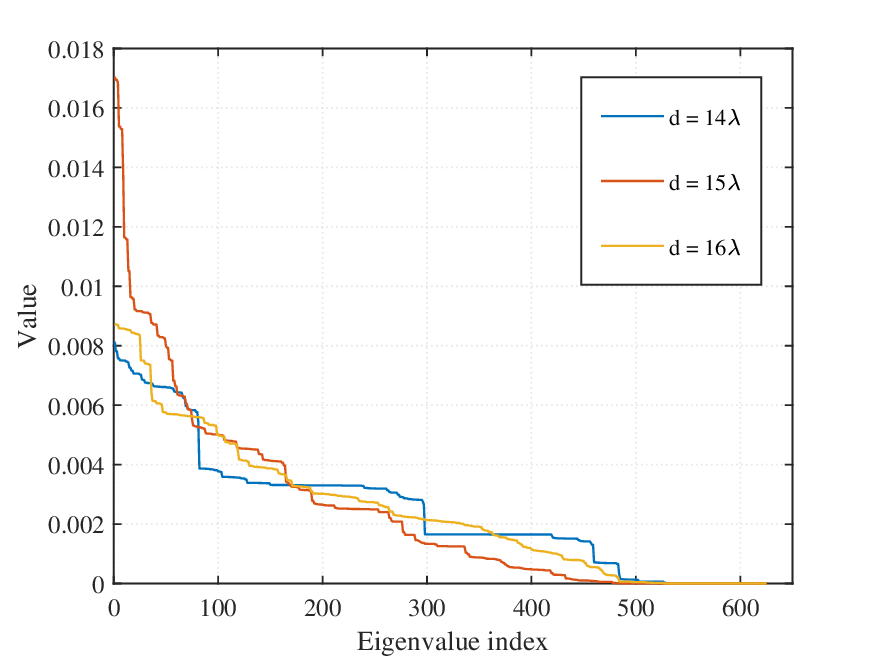}\\  \vspace{-0.2cm}
  \caption{Eigenvalues after the threshold}\label{p8} \vspace{-0.4cm}
\end{figure}

The above result can be explained by the paraxial situation aforementioned in Section II.
Before the threshold, the XL-MIMO system works in paraxial situation.
As shown in Fig. \ref{p7}, the significant eigenvalues of the correlation matrix are nearly constant,
and an abrupt drop appears as eigenvalue index number increases.
In Remark \ref{Re1}, we have given a brief mathematical proof that 
$n_{\mathrm{EDoF2}}$ can well approximate the $n_{\mathrm{EDoF}}$ in paraxial situation.
After the threshold, the XL-MIMO system no longer works in paraxial situation.
In Fig. \ref{p8}, a continuous decrease of eigenvalues with their index number can be observed,
which makes the estimation methods no longer effective.
Therefore, the derived threshold can be regarded as a boundary between the paraxial and non-paraxial situations.

\begin{figure}[t]
\vspace{-0.8cm}
  \centering
  \includegraphics[scale=0.4]{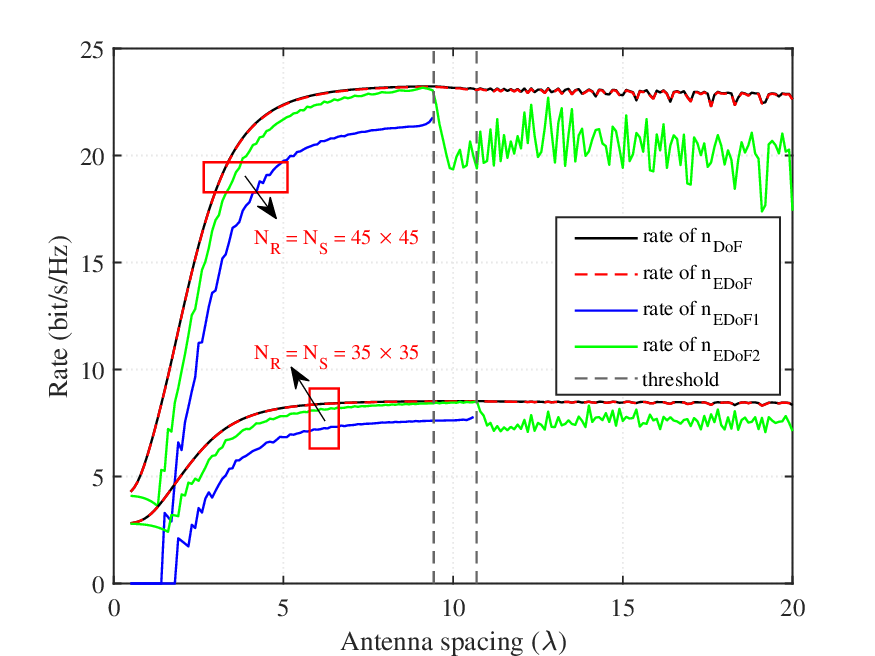}\\  \vspace{-0.2cm}
  \caption{Channel capacity with EDoF}\label{p9} \vspace{-0.4cm}
\end{figure}

Fig.~\ref{p9} illustrates the channel capacities obtained by $n_{\mathrm{DoF}}$, $n_{\mathrm{EDoF}}$, $n_{\mathrm{EDoF1}}$ and $n_{\mathrm{EDoF2}}$.
It is readily seen that the rates of $n_{\mathrm{DoF}}$ and $n_{\mathrm{EDoF}}$ almost match perfectly.
In the range smaller than the threshold of antenna spacing, 
the rate of $n_{\mathrm{EDoF2}}$ matches well with the rate of $n_{\mathrm{EDoF}}$,
which verifies the effectiveness of the estimation method in \eqref{Req4}.
Although a gap exists between the rates of $n_{\mathrm{EDoF1}}$ and $n_{\mathrm{EDoF}}$,
the two rates show a similar growth trend,
which means that the method in \eqref{Req3} can be regarded as a rough estimation on $n_{\mathrm{EDoF}}$.
When the antenna spacing is over the threshold, 
we can see a randomly varying gap between the rates of $n_{\mathrm{EDoF2}}$ and $n_{\mathrm{EDoF}}$,
which confirms that the estimation method in \eqref{Req4} is no longer effective.

\section{Conclusion} 

We investigated a near-field XL-MIMO communication system.
Based on the Green's function-based channel model, we discussed the channel capacity, which is related to the $n_{\mathrm{EDoF}}$. 
We further analyzed that two widely used EDoF estimation methods are effective in paraxial situation.
To increase $n_{\mathrm{EDoF}}$, we focused on the impact of antenna spacing.
Then, from the perspective of the array gain, an approximate closed-form expression for the threshold of antenna spacing was derived when $n_{\mathrm{EDoF}}$ was maximized.
The numerical results verified the accuracy of the main results of this letter.
Furthermore, the derived threshold was found to be the boundary between the paraxial and non-paraxial situations,
which means that the EDoF estimation methods are effective in the range smaller than the threshold of antenna spacing.

\begin{appendices}

\section{}\label{app_A}

When the transmit array focuses on $\mathbf{r}_0$, 
the received signal at antenna $\mathbf{r}_1$ can be expressed as \vspace{-0.1cm}
\begin{equation}\label{Appeq1} \vspace{-0.1cm}
  f_1=\sum_{n=1}^{\sqrt{N}}{\sum_{m=1}^{\sqrt{N}}{t_{n,m}^{1}e^{i\theta _{n,m}^{0}}s}}+n ,
\end{equation}
where $t_{n,m}^{1}$ is the channel coefficient for antenna $\mathbf{r}_1$, given by \vspace{-0.1cm}
\begin{equation}\label{Appeq2}  \vspace{-0.1cm}
   t_{n,m}^{1} \approx -\frac{e^{i\frac{2\pi}{\lambda}\sqrt{\bar{x}_{n}^{2}+\bar{y}_{m}^{2}+L^2+d^2-2\bar{x}_nd}}}{4\pi L} .
\end{equation}
Thus, the SNR at antenna $\mathbf{r}_1$ can be expressed as \vspace{-0.1cm}
\begin{equation}\label{Appeq3} \vspace{-0.1cm}
  \mathrm{SNR}_1=\frac{P}{N\sigma _{\mathrm{n}}^{2}}\left| \sum_{n=1}^{\sqrt{N}}{\sum_{m=1}^{\sqrt{N}}{t_{n,m}^{1}e^{i\theta _{n,m}^{0}}}} \right|^2=\frac{P}{\sigma _{\mathrm{n}}^{2}}\frac{1}{\left( 4\pi L \right) ^2}\rho _1 ,
\end{equation}
with the array gain $\rho _1$ expressed as \vspace{-0.1cm}
\begin{equation}\label{Appeq4} \vspace{-0.1cm}
  \rho _1=\frac{1}{N}\left| \sum_{n=1}^{\sqrt{N}}{\sum_{m=1}^{\sqrt{N}}{e^{i\frac{2\pi}{\lambda}\sqrt{\bar{x}_{n}^{2}+\bar{y}_{m}^{2}+L^2+d^2-2\bar{x}_nd}}e^{i\theta _{n,m}^{0}}}} \right|^2.
\end{equation}

Then, substituting \eqref{ASeq6} into \eqref{Appeq4}, we further calculate the array gain $\rho _1$ as follows \vspace{-0.2cm}
\begin{equation*}  \vspace{-0.2cm}
  \rho _1
  \!=\!  \frac{1}{N} \! \left| \sum_{n=1}^{\sqrt{N}} \!
  {\sum_{m=1}^{\sqrt{N}} \! {e^{i\frac{2\pi}{\lambda} \! \sqrt{\bar{x}_{n}^{2}+\bar{y}_{m}^{2}+L^2+d^2-2\bar{x}_nd}}e^{-i\frac{2\pi}{\lambda} \! \sqrt{\bar{x}_{n}^{2}+\bar{y}_{m}^{2}+L^2}}}} \right|^2
\end{equation*}
\begin{equation*} \vspace{-0.2cm}
  \approx \! \frac{1}{N} \! \left| \sum_{n=1}^{\sqrt{N}} \! {\sum_{m=1}^{\sqrt{N}} \! {e^{i\frac{2\pi}{\lambda} \! \left( L+\frac{\bar{x}_{n}^{2}}{2L}+\frac{\bar{y}_{m}^{2}}{2L}+\frac{d^2-2\bar{x}_nd}{2L} \right)}e^{-i\frac{2\pi}{\lambda} \! \left( L+\frac{\bar{x}_{n}^{2}}{2L}+\frac{\bar{y}_{m}^{2}}{2L} \right)}}} \right|^2
\end{equation*}
\begin{equation*} \vspace{-0.2cm}
  = \! \left| \sum_{n=1}^{\sqrt{N}}{e^{-i\frac{2\pi}{\lambda}\left( \frac{d^2-2\bar{x}_nd}{2L} \right)}} \right|^2
  \overset{\left( a \right)}{=} \left| \sum_{n=1}^{\sqrt{N}}{e^{-i\frac{2\pi}{\lambda}\left( \frac{d}{2}-\left( n-\frac{\sqrt{N}+1}{2} \right) d \right) \frac{d}{L}}} \right|^2
\end{equation*}
\begin{equation}\label{Appeq5} \vspace{-0.2cm}
  \hspace{-4.5cm}
  = \! \left| \sum_{n=1}^{\sqrt{N}}{e^{-i\frac{2\pi}{\lambda}\left( n-\frac{\sqrt{N}+2}{2} \right) \frac{d^2}{L}}} \right|^2.
\end{equation}
The approximation in \eqref{Appeq5} is based on the assumption of $L^2\gg \bar{x}_{n}^{2},\bar{y}_{m}^{2}$ and the Taylor approximation $\sqrt{1+x}\approx 1+\frac{x}{2}$ as $x \rightarrow 0$.
Step $\left( a \right)$ is obtained by substituting \eqref{ASeq3} into the expression.
Then, the summation term can be derived as \vspace{-0.2cm}
\begin{equation*} \vspace{-0.3cm}
  \sum_{n=1}^{\sqrt{N}}{e^{-i\frac{2\pi}{\lambda}\left( n-\frac{\sqrt{N}+2}{2} \right) \frac{d^2}{L}}}=\frac{e^{i\frac{2\pi}{\lambda}\frac{\sqrt{N}}{2}\frac{d^2}{L}}\left( 1-e^{-i\frac{2\pi}{\lambda}\sqrt{N}\frac{d^2}{L}} \right)}{1-e^{-i\frac{2\pi}{\lambda}\frac{d^2}{L}}}
\end{equation*}
\begin{equation*}  \vspace{-0.15cm}
  =\frac{e^{i\frac{2\pi}{\lambda}\frac{\sqrt{N}}{2}\frac{d^2}{L}}-e^{-i\frac{2\pi}{\lambda}\frac{\sqrt{N}}{2}\frac{d^2}{L}}}{e^{-i\frac{\pi}{\lambda}\frac{d^2}{L}}\left( e^{i\frac{\pi}{\lambda}\frac{d^2}{L}}-e^{-i\frac{\pi}{\lambda}\frac{d^2}{L}} \right)}=e^{i\frac{\pi}{\lambda}\frac{d^2}{L}}\frac{2i\sin \left( \frac{\sqrt{N}\pi d^2}{\lambda L} \right)}{2i\sin \left( \frac{\pi d^2}{\lambda L} \right)}
\end{equation*}
\begin{equation}\label{Appeq6}
  \hspace{-4.1cm}
  =e^{i\frac{\pi d^2}{\lambda L}}\frac{\sqrt{N}\sin\mathrm{c}\left( \frac{\sqrt{N}d^2}{\lambda L} \right)}{\sin\mathrm{c}\left( \frac{d^2}{\lambda L} \right)}.
\end{equation}

Substituting \eqref{Appeq6} into \eqref{Appeq5}, we finish the derivations for \eqref{T1eq1}.

\end{appendices}

\bibliographystyle{IEEEtran}
\bibliography{IEEEabrv,Reference}

\end{document}